\documentstyle[aps,epsfig]{revtex}

\newcommand{\be}{\begin{equation}}
\newcommand{\ee}{\end{equation}}
\newcommand{\bc}{\begin{center}}
\newcommand{\ec}{\end{center}}
\newcommand{\bi}{\begin{itemize}}
\newcommand{\ei}{\end{itemize}}
\newcommand{\ba}{\begin{eqnarray}}
\newcommand{\ea}{\end{eqnarray}}
\newcommand{\ie}{{\it i.e.\ }}

\newcommand{\ignore}[1]{}

\begin{document}
\twocolumn[\hsize\textwidth\columnwidth\hsize\csname
@twocolumnfalse\endcsname

\title{Highly clustered scale-free networks}

\author{Konstantin Klemm\cite{kk} and V\'{\i}ctor M. Egu\'{\i}luz\cite{vme}
\\Center for Chaos and Turbulence Studies\cite{cats}
\\ Niels Bohr Institute, Blegdamsvej 17, DK-2100 Copenhagen \O, Denmark}

\date{\today}

\maketitle


\ignore{
\renewcommand{\baselinestretch}{2}
\begin{document}
\begin{center}
Classification: Physical Sciences --- Applied Physical Sciences

\vspace*{.8cm}
{\huge Highly clustered scale-free networks}

\vspace*{1cm}
{\large Konstantin Klemm and V\'{\i}ctor M. Egu\'{\i}luz
\\Center for Chaos and Turbulence Studies
\\Niels Bohr Institute, Blegdamsvej 17, DK-2100 Copenhagen \O, Denmark
}
\end{center}
\vspace*{5mm}
Corresponding author: Konstantin Klemm, Center for Chaos and Turbulence
Studies, Niels Bohr Institute, Blegdamsvej 17, DK-2100 Copenhagen \O,
Denmark
Tel. +45\ 35\ 32\ 52\ 73, Fax +45\ 35\ 32\ 54\ 25, e-mail
{\tt klemm@nbi.dk}

\vspace*{5mm}
\noindent This manuscript contains 28 pages (including this front page),
4 figures and no tables. The length of the abstract is 128 words. The
character count is 43537. 

\vspace*{5mm}\noindent Non-standard abbreviation: ``BA-model''
\clearpage
}

\begin{abstract}
We propose a model for growing networks based
on a finite memory of the nodes. The model shows stylized features of
real-world networks: power law distribution of degree, linear preferential
attachment of new links and a negative correlation between the age of a node
and its link attachment rate. Notably, the degree distribution is conserved
even though only the most recently grown part of the network is considered.
This feature is relevant because real-world networks truncated
in the same way exhibit a power-law distribution in the degree.
As the network grows, the clustering reaches an asymptotic value larger than
for regular lattices of the same average connectivity. These high-clustering
scale-free networks indicate that memory effects could be crucial for a correct
description of the dynamics of growing networks.
\end{abstract}
]

\noindent Many systems can be represented by networks, \ie as a set of nodes
joined together by links. Social networks,
the Internet, food webs, distribution networks,
metabolic and protein networks, the networks of airline routes, scientific
collaboration networks and citation networks are just some examples of such systems
\cite{Strogatz01,Amaral00,Wasserman94,Watts98,Albert99,Williams00,Jeong00,Jeong01a,Redner98,Newman01a,Barabasi99}.
Recently it has been observed that a variety of networks exhibit topological
properties that deviate from those predicted by random graphs
\cite{Strogatz01,Amaral00}. For instance, real networks display
{\em clustering} higher than expected for random networks \cite{Watts98}.
Also, it has been found that many large networks are {\em scale-free}.
Their degree distribution decays as a power-law that cannot be accounted for
by the Poisson distribution of random graphs \cite{Erdos60,Bollobas85}.
The type of the degree distribution is of
great importance for the functionality of the network
\cite{Albert00,Cohen00,Cohen01}. Beside
the degree distribution, other features of the growth dynamics of real-world
networks are currently under investigation. For citation networks, the
Internet, and collaboration networks of scientists and actors, it has been
shown \cite{Jeong01b,Newman01b} that the probability for a node to obtain a
new link is an
increasing function of the number of links the node already has. This feature
of the dynamics is called {\em preferential attachment}. Furthermore the
aging of nodes is of particular interest \cite{Dorogovtsev00a}. In the
network of scientific collaborations, every node stops receiving links
a finite time after it has been added to the network, since scientists have
a finite time span of being active. Similarly, in citation networks,
papers cease to receive links (citations), because their contents are
outdated or summarized in review articles, which are then cited instead.
Whether a paper is still cited or not, depends on a collective
{\em memory} containing the popularity of the paper.

In the current paper we address the study of growing complex networks from
the perspective of the memory of the nodes. First, we present empirical
evidence for the age dependence of
the growth dynamics of the network of scientific citations. We find that old
nodes are less likely to obtain links than nodes added
to the network more recently. Second, motivated by this finding, we
introduce a model of network
self-organization that accounts for the three empirical features mentioned
before: (1) power law distribution for the degree, (2) preferential attachment,
and (3) negative correlation between age and attachment rate. The clustering
of the generated networks is higher than in corresponding regular lattices,
justifying the name {\em highly clustered scale-free networks}.

\section*{Previous models}

The earliest and most basic model generating scale-free networks
has been introduced by Barab\'asi and Albert \cite{Barabasi99},
henceforth we use the acronym BA-model. This model explicitly
incorporates the preferential attachment in the dynamical rules.
At each time step a new node is added to the network and new links
are attached from this new node to old nodes. The probability
that a node obtains an additional link is proportional to its
current degree. It can be interpreted as an
application of Simon's growth model in the context of networks
\cite{Simon55,Bornholdt00}, readily explaining the emergent
scaling in the degree distribution.
The BA-model has been successively modified reproducing the scale-free
behavior of the connectivity distribution
\cite{Albert00b,Krapivsky00,Dorogovtsev00b}. For the sake of clarity,
in the remaining of the paper
we will refer to the BA-model as a well-established model of
growing scale-free networks.

\begin{figure}
\centerline{\epsfig{file=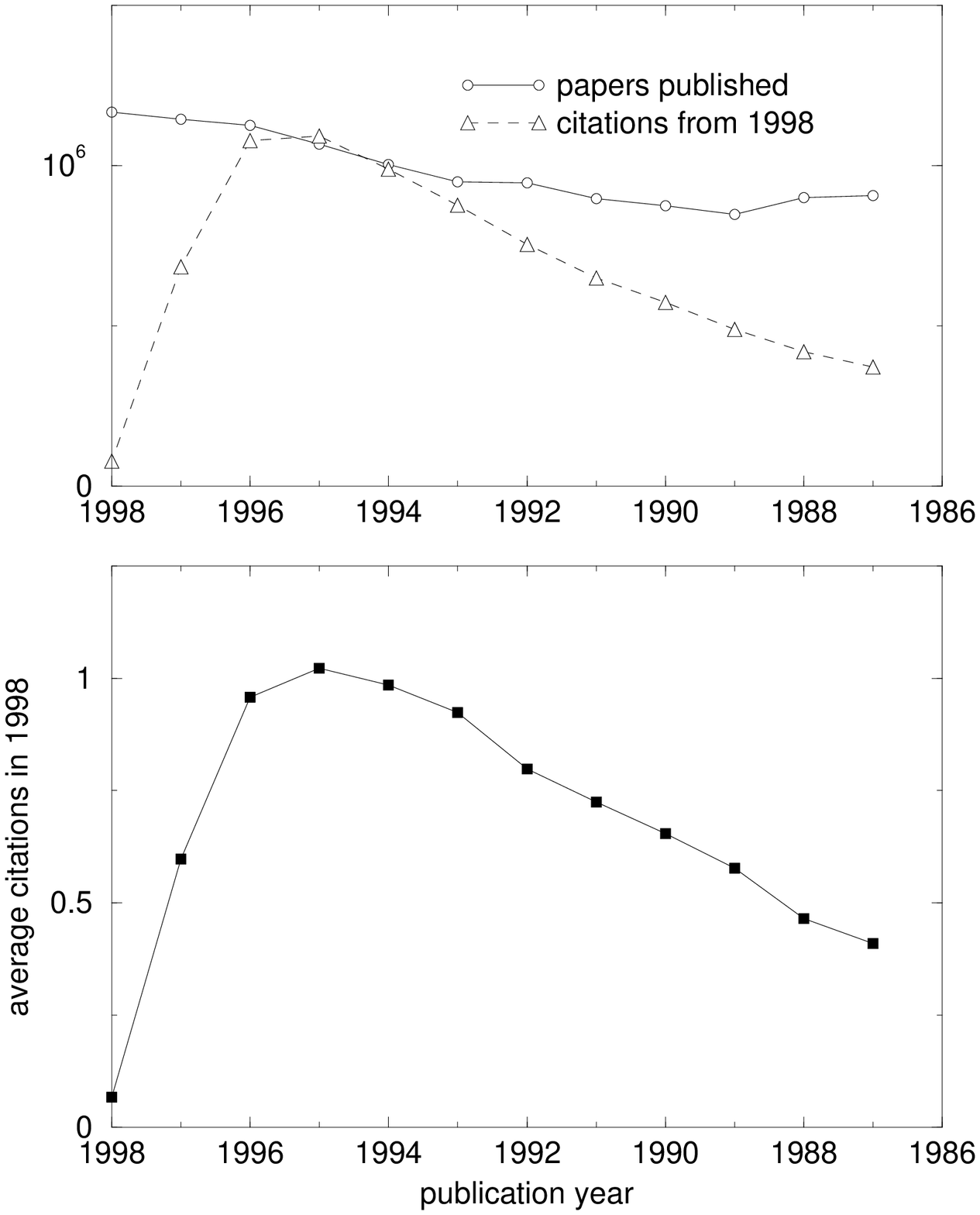,width=.5\textwidth}}
\caption{Data on the network formed by scientific publications (nodes)
and citations (directed links).
Upper panel, circles: The number of papers published in a given year
from 1987 to 1998.
Triangles: The total number of citations made
in papers published in 1998 and referring to papers published in a given
year \protect\cite{Raan00}. The data for both curves have been
extracted from the ISI database \protect\cite{ISI}.
Lower panel: The average number of citations (incoming links) a paper received
in 1998 as a function of the paper's publication year. The values are obtained
as the ratio between the values of the two curves in the
upper panel. Considering only papers more than 3 years old (published before
1995) the rate of obtaining new citations decreases with age. This indicates
that aging is an important feature of citation networks.}
\label{fcitation}
\end{figure}

Real-world networks have properties that cannot be accounted for 
by the BA-model. We find a discrepancy with
respect to empirical data in the
correlation between a node's age and its rate of acquiring links.
For the network of scientific citations this correlation is
negative: the mean rate of citations a paper receives decreases
with increasing age. This is supported by citation rate data of
the years 1987-1998, shown in Figure \ref{fcitation}. Except for the three
first years prior to the publication year, the citation rate decreases with
age \cite{Raan00}. In contradiction to this empirical result, in the BA-model
the mean
attachment rate is positively correlated with age. Here the
attachment rate is proportional to the degree, being largest for
the oldest nodes since these began accumulating links earliest. A
further consequence of this feature is a strong positive
correlation between the age of a node and its degree. This kind of
correlation has not been found in the network formed by the
hyperlinks of the World Wide Web \cite{Adamic00}. We also notice that
if the oldest
nodes are disregarded, the networks generated by the BA-model are
not scale-free any more. However, real-world networks have shown
to be scale-free even though they are truncated, \ie the major
part of the oldest nodes is disregarded.

\section*{Growth and deactivation model}

The shortcomings indicated in the previous paragraph motivate our attempt to
model
self-organization of scale-free networks. The approach presented here is based
on the degree-dependent deactivation dynamics of the nodes. Preferential
attachment and the convergence to a power-law degree distribution are shown to
be emergent properties of the dynamics.

The model describes the growth dynamics of a network with directed links. By
$k_i$ we denote the in-degree of node $i$, \ie the number of links pointing to
node $i$. Each node of the network can be in two different states: active or
inactive. A new node added to the network is always in the {\em
active} state first. It receives links from subsequently generated nodes
until it is deactivated. Then the node does not receive links any more.
The transition of a node from the active to the inactive state can be
interpreted as a collective ``forgetting'' of the node since new nodes do not
connect
to it any more. For the construction of the model we assume that the
probability rate $P$ of deactivation decreases with the in-degree of the
node.
Considering for instance the case of citation networks, this means that the
more often a paper has been cited, the less likely it is forgotten.
Specifically, we make the assumption that the deactivation probability can be
written as
$P\propto (k+a)^{-1}$, where $a>0$ is a constant bias.

At any step of the time-discrete dynamics $m$ nodes in
the network are active, all the other nodes are inactive. As the initial
condition we use a network consisting of $m$ active, completely connected
nodes. Then the dynamics runs as follows:
\begin{enumerate}
\item Add a new node $i$ to the network. The new node is disconnected at
first, so $k_i=0$ at this point.
\item Attach $m$ outgoing links to the new node $i$. Each node $j$ of the
 $m$ active nodes receives exactly one incoming link, thereby
 $k_j \rightarrow k_j+1$.
\item Activate the new node $i$.
\item \label{dyndef}
Deactivate one of the active nodes. The probability that the node $j$
 is deactivated is given by
 \be
 P(k_j) = \frac{\gamma - 1}{a + k_j}~,
 \label{eqdeactp}
 \ee
 where $a>0$ is a constant bias and the normalization factor
 is defined as $\gamma - 1= \left(
 \sum_{l\in\cal A} \frac{1}{a + k_l} \right)^{-1}$. The summation runs
 over the set ${\cal A}$ of the currently active nodes.
\item Resume at 1.
\end{enumerate}
The average connectivity of the network is given by the
number of outgoing links per node, $m$. It is worth noting that a node
receives incoming
links during the lifetime $T$ it is active, and once inactive it will not
receive links any longer. Thus for each node $i$ the time $T_i$ spent in the
active state and the in-degree $k_i$ are equivalent.

The deactivation mechanism strongly simplifies the dynamics of growing complex
networks. Neither gradual aging nor possible reactivation are taken into
account. For instance, in the context of citation networks, the model
does not consider the rediscovery of ``forgotten'' papers. Moreover,
the functional form of the deactivation probability might well differ from
Eq.\ (\ref{eqdeactp}). However, we will show that the model reproduces
several features of real growing networks.

\section*{Degree distribution}

The distribution $N(k)$ of the in-degree $k$ can be obtained analytically for
the model defined above, considering the continuous limit of $k$.
Let us first derive the distribution $p^{(t)}(k)$ of the in-degree of the
active nodes at time $t$. For $k>0$, the time evolution is determined by the
following master equation
\ba \label{masterp}
p^{(t+1)}(k+1) & = & \left( 1- P(k) \right) p^{(t)}(k) \nonumber \\
               & = & \left(1-\frac{\gamma-1}{a+k} \right)p^{(t)}(k)
\label{eqP}
\ea
where $a$ and $\gamma$ are defined in step \ref{dyndef} of the model
definition. The boundary value p(0) is a constant reflecting the constant
rate of new nodes with initial $k=0$.

Assuming that the fluctuations of the normalization $\gamma-1$ are small
enough, such that $\gamma$ may be treated as a constant, the stationary case
$p^{(t+1)}(k)=p^{(t)}(k)$ of Eq.~(\ref{masterp}) yields
\be
p(k+1)-p(k)=-\frac{\gamma-1}{a+k} p(k)~.
\label{eqProb}
\ee
Treating $k$ as continuous we write
\be
\frac{{\rm d}p}{{\rm d}k} = - \frac{\gamma-1}{a + k} p(k)~,
\ee
and obtain the solution
\be
p(k) = b (a + k)^{-\gamma+1}~,
\label{eqPk}
\ee
with appropriate normalization constant $b$.
In case the total number $n$ of nodes in the network is large
compared with the number $m$ of active nodes, the overall degree distribution
$N(k)$ can be approximated by considering the inactive nodes only. Thus $N(k)$
can be calculated as the rate of change of the degree distribution $p(k)$ of the active
nodes. We find
\be
N(k) = - \frac{{\rm d}p}{{\rm d}k} =  c (a+k)^{-\gamma}
\label{eqN}
\ee
with $c = (\gamma-1) a^{\gamma-1}$.
The exponent $\gamma$ is obtained from a self-consistency condition
obtained from the average connectivity
\be
m = c \int_0^\infty \frac{k}{(a+k)^\gamma}  dk~,
\ee
which gives
\be \label{eqgamma}
\gamma = 2 + \frac{a}{m}~.
\label{gamma}
\ee
Thus the exponent $\gamma$ depends only on the ratio $a/m$.
Similar expressions 
have been obtained for a version of the BA-model with directed links
\cite{Bornholdt00,Dorogovtsev00b}. 
Although the growth and deactivation model has been formulated
for directed networks, it can be
easily applied also to generate undirected networks.

\begin{figure}
\centerline{\epsfig{file=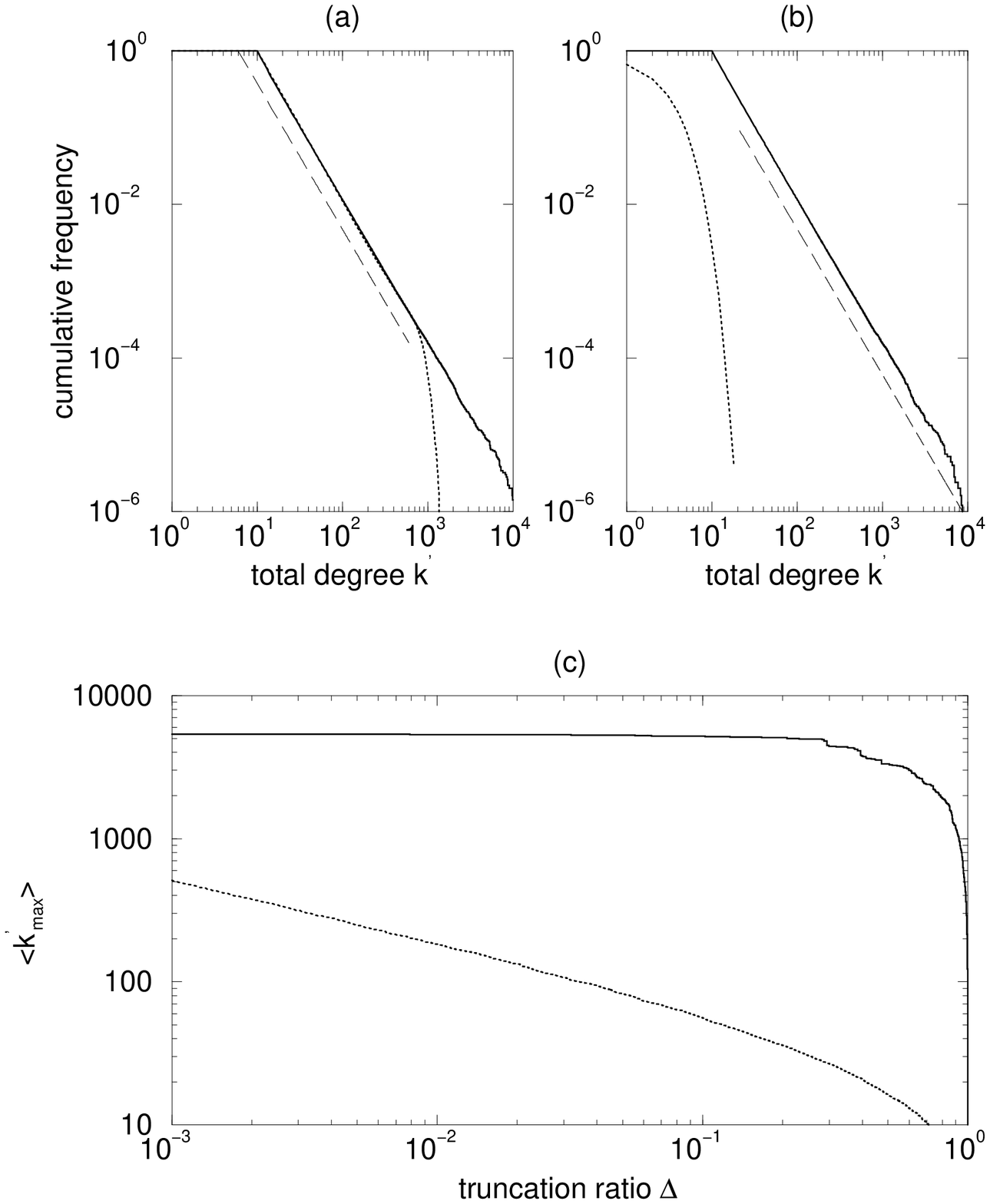,width=.5\textwidth}}
\caption{Comparison of the degree distribution obtained for the
undirected networks following the BA (dashed line)
and the growth and deactivation model (solid line). In (a) the complete
networks are considered after
$5\times 10^4$ time steps. In contrast,
in (b) only the network formed by the  newest nodes and their links is taken into account.
In (c) we plot the maximum degree, $k_{max}$, observed in the truncated network
against the truncation ratio $\Delta$.
In the
BA model, $k_{max}$ scales as a power law with $\Delta$.
However, the degree distribution in the new model shows
a power law distribution of degree, whose cutoff is only
slightly affected by the finite size of the truncated network.
All curves are averages over 100 independent simulation runs.
}
\label{fdegree}
\end{figure}

\subsection*{Numerical results}

Figure~\ref{fdegree}(a) shows the cumulative distribution of the
total degree $k^\prime = (m+k)$ obtained by simulating the model
for $5\times 10^4$ time steps. We obtain a power law scaling for
several decades, in agreement with the analytical result in
Eq.~\ref{eqN}. The exponent found numerically is 1.9, slightly
below the analytical result $\gamma -1 = 2+ a/m -1 = 2$ for the
case $a=m$. The deviation can be explained by the continuous limit
used in the theoretical derivation of $\gamma$ and the assumption that
$\gamma$ is a constant. Conducting further simulations for various
values of $m$ and $a$, we find that the fluctuations of $\gamma$
become smaller when increasing $m$ and/or $a$. Then the
discrepancy between analytical and numerical results decreases.
Figure~\ref{fdegree}(a) also shows corresponding simulation results
for the BA model, using $m=10$ and $5\times 10^4$ time steps as
well. In the range $k^\prime<1000$ we obtain almost the same
distribution as for the growth and deactivation model.
However, the main difference between both models is the presence of a cutoff at
a lower value for the BA-model.

Up to this point we have considered degree distributions including {\em all}
nodes of the network. However, in many cases empirical data contain only those
nodes and links of the network that have been created most recently. For
instance, studies on scientific citation networks \cite{Redner98} are
restricted to papers that are not older than 20 years, thereby ignoring the
major part of the initial network. A pronounced power law regime is observed in
the degree distribution of these {\em truncated} networks. Therefore it is
important to investigate the robustness of the scale-free networks obtained
from models under truncation in time. Figure~\ref{fdegree}(b) shows the
cumulative degree distributions analogous to Fig.~\ref{fdegree}(a), but now
regarding the truncated network where the fraction $\Delta=50\%$ of oldest
nodes and all their links are disregarded.
Concerning the BA-model the effect of truncation is drastic. The truncated
network does not exhibit a scale-free range in the degree distribution. This is
different for the growth and deactivation model. The influence of the
truncation on the degree distribution is a slight shift of the cutoff for high
$k^\prime$. In order to view systematically the effect of truncation, we
consider the largest degree $k_{max}^\prime$, occurring in the truncated
network, as a function of the fraction $\Delta$ of disregarded nodes. According
to Fig.~\ref{fdegree}(c), $k_{max}^\prime$ decays as a power law (with an
approximate exponent of 0.5, $k_{max}^\prime \sim \Delta^{-0.5}$) for the
BA-model. On the other hand, the new model introduced here exhibits only a weak
dependence of the maximum degree on the truncation.

\section*{Linear preferential attachment}

Another relevant dynamical property is the degree-dependent attachment rate
$\Pi(k)$. It is measured as follows: Consider the set ${\cal K}$ of nodes with
degree $k$ at a certain time $t$. Measure the average degree $k+\Delta k$ of
the nodes in ${\cal K}$ at a later time $t+\Delta t$. Then let $\Pi(k) =
\Delta k / \Delta t$. In recent studies of various growing networks, it has
been found empirically that $\Pi(k)$ is an increasing function
\cite{Jeong01b,Newman01b,Barabasi01}. This phenomenon is called preferential attachment.
For the Internet and citation networks the preferential attachment is linear,
$\Pi(k)\propto k$.

We can calculate $\Pi(k)$ for the model introduced in the present
Paper. At a time $t$, the network contains $t$ nodes. $tN(k)$ of these have
degree $k$. The number of active nodes with degree $k$ is $mp(k)$.
A time step later, $\Delta t=1$, each of the active nodes has increased its
degree by 1, whereas the degree of the inactive nodes remains unchanged.
Thus, according to Eqs.\ (\ref{eqPk}) and (\ref{eqN}),
the average increase of the degree is
\be
\Pi(k) = \frac{mp(k)}{tN(k)} \propto (a + k)~.
\label{eqPi}
\ee
The model shows linear preferential attachment as an emergent
property of the degree-dependent deactivation dynamics.

\section*{Age distribution}

\begin{figure}
\centerline{\epsfig{file=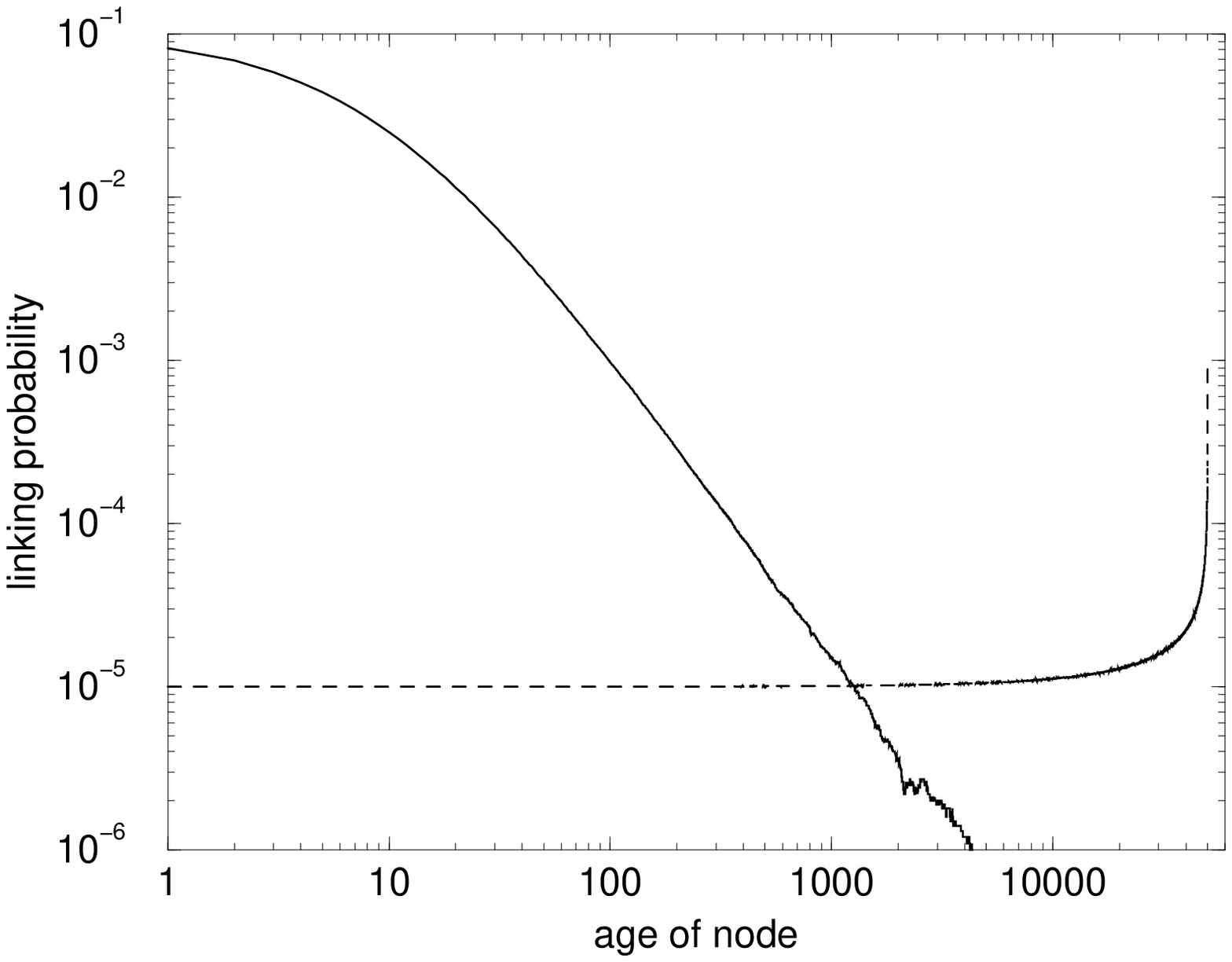,width=.5\textwidth}}
\caption{Age distribution $h(\tau, t)$ of nodes receiving links.
In the growth and deactivation model the distribution (solid line)
follows a power law decay with the age of the node. In contrast,
in the BA-model (dashed line) it is the oldest nodes that are most likely
to receive new links. For each of the two models the plotted data have been
generated as an average over 100 independent simulation runs lasting
$5 \times 10^4$ time steps.}
\label{fage}
\end{figure}

Let us now consider the distribution of the age $\tau$ of nodes receiving
a new link. We define the time-dependent age distribution
$h(\tau,t)$ as the probability that a new link created at time $t$ attaches
to a node of age $\tau$, \ie to a node created at time $t-\tau$.
For the model defined here, the age distribution $h$ is easy to obtain.
Only active nodes receive links, and for these nodes their age $\tau$
and their in-degree $k$ have the same value. Therefore the probability that
the node of age $\tau$ obtains a new link is the same as the probability for
a node with $\tau$ links to be active, given by equation (\ref{eqPk}). It is
independent of $t$:
\be \label{eqagegd}
h(\tau) \propto (a + \tau)^{-\gamma+1}~.
\ee
For comparison we calculate the age distribution for the BA-model. Apart from
small deviations, the
total degree of the node $i$ created at time $t_i$ is \cite{Barabasi99}:
\be
k^\prime_i = m \left( \frac{t}{t_i}\right)^{0.5}
           = m \left( \frac{t}{t-\tau}\right)^{0.5}~,
\ee
where the second equality is due to the substitution $t_i=\tau-t$.
The probability of obtaining a new link is proportional
to the total degree, thus we find
\be
h(\tau,t) = \frac{1}{2mt} m \left(\frac{t}{t-\tau}\right)^{0.5}
            = \frac{1}{2} [t(t-\tau)]^{-0.5}~.
\ee
In the BA-model the probability of receiving a new link increases with the
age of the node. In sharp contrast, the growth and deactivation model
displays a forgetting of old nodes where
the rate of forgetting is a power-law, Eq.\ (\ref{eqagegd}). Figure \ref{fage}
shows plots of the age distributions for both models, to be compared with the
empirical data in Fig. \ref{fcitation}. The age distribution
of the growth and deactivation model decays with $\tau$. This agrees with the
empirical data on citation networks except for the first 3 years after
publication.

\section*{Clustering coefficient}

The clustering coefficient $C$ \cite{Watts98} is one of the parameters
used to characterize the topology of complex networks. It is a
local property measuring the probability with which two neighbors
of a node are also neighbors to each other (nodes $i$ and $j$ are
neighbors if there is a link between $i$ and $j$). It has been
found that many real world networks present a clustering
coefficient much larger than the corresponding random graph, which
scales with the system size $N$ as $C_{rand} \sim \langle k
\rangle / N$.

\begin{figure}
\centerline{\epsfig{file=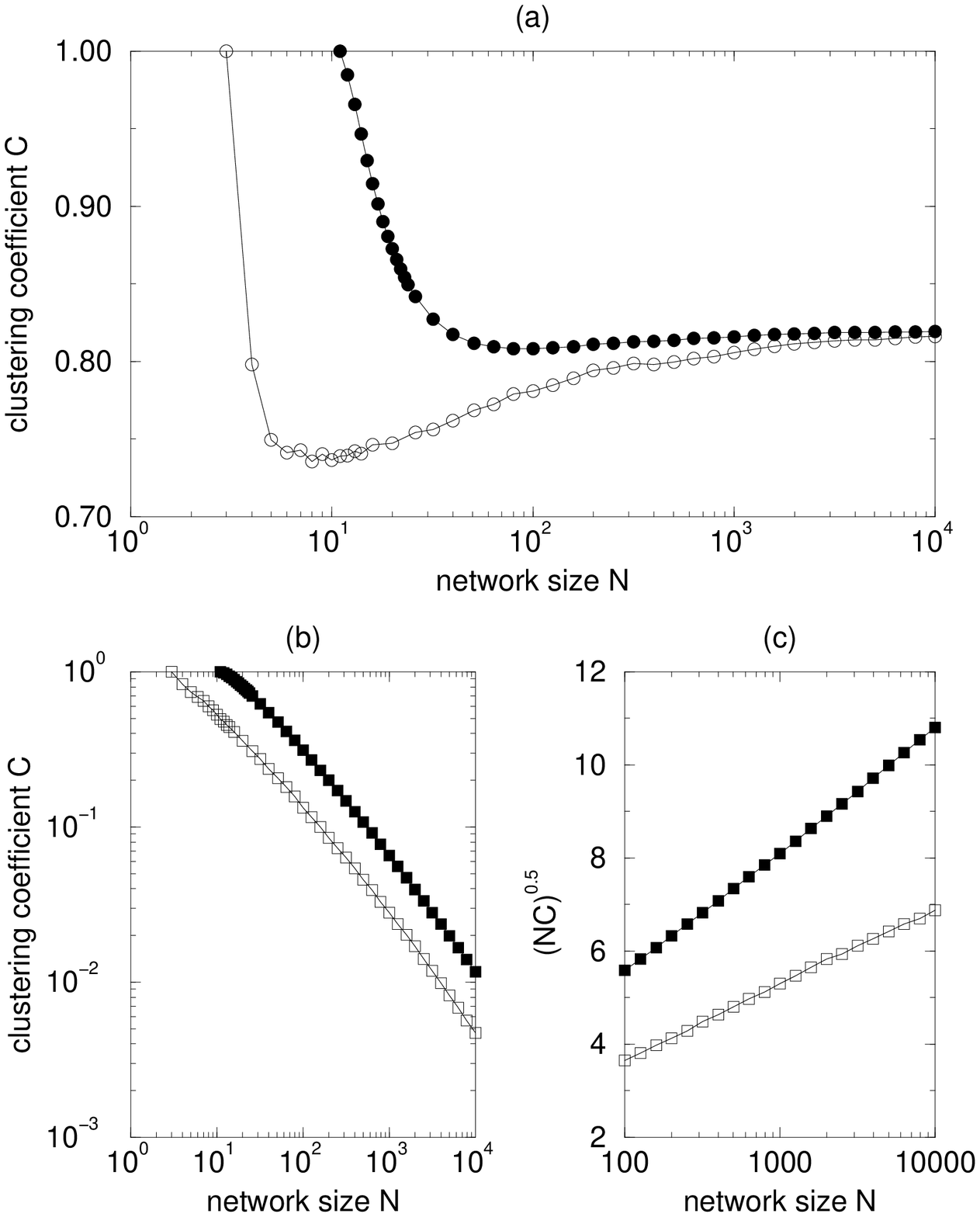,width=.5\textwidth}}
\caption{Dependence of the clustering coefficient $C$ on the size $N$ of the
network. (a) Growth and deactivation model for $m=a=2$ (unfilled) and $m=a=10$
(filled symbols). $C$ approaches a high stationary value above 0.8. Note
that corresponding one-dimensional regular lattices have $C=0.5$ ($m=2$) and
$C=0.71$ ($m=10$) respectively (b) BA-model for $m=2$ (unfilled) and $m=10$
(filled symbols). The clustering coefficient strongly decreases as the network
grows. (c) The same data as in (b), but plotting $(NC)^{0.5}$ as a function of
$N$. This function is a straight line in a log-linear plot, indicating that
$C$ scales as $(\ln N)^2/N$ for large $N$. Each data point is an average over
100 independent simulation runs. The clustering coefficient
\protect \cite{Watts98} is defined as follows: Consider a node $i$ with total
degree $k^\prime_i$. Between the $k^\prime_i$ nodes that
$i$ is linked with, at most $k(k-1)/2$ links are possible. Let $C_i$ denote
the fraction of links that actually exist among the neighbors of $i$.
The clustering coefficient $C$ is the average of $c_i$ taken over all $N$
nodes $i$ in the network. Note that all links are considered as bidirectional
when calculating the clustering coefficient.}
\label{fclust}
\end{figure}

Fig.~\ref{fclust}(a) shows that for the growth and deactivation model the
clustering coefficient tends towards an asymptotic value ($\approx0.83$) as
the network grows. The analytical derivation of $C$ is facilitated by the
observation, that the clustering $C_i$ of a node merely depends on the node's
in-degree $k_i$. A detailed calculation gives an asymptotic value $C = 5/6$
for the case of $a=m$ considered here. Thus the model generates networks with
a higher clustering than the corresponding one-dimensional regular lattices,
$C_{1D} < 3/4$. The large value of the clustering
coefficient and the fact that it does not decrease with network size is
in qualitative agreement with recent data on the Internet \cite{Pastor01}.
For the sake of comparison, in Fig.~\ref{fclust}(b) the clustering coefficient
of the BA-model is plotted for several network sizes $N$.
Here the clustering clearly decays with increasing $N$.
The quantitative behavior of the decay can be described by
$C \sim (\ln N)^2/N$. The detailed derivation of the clustering coefficients
for both models is included in (Klemm and Egu\'{\i}luz, unpublished work). 

\bigskip
\section*{Conclusions}

The analysis of citation networks suggests a negative correlation between the
age of a node and its probability to obtain further links. Older nodes are less
likely to increase their connectivity than those added to the network more
recently. Motivated by this finding, we have proposed and analyzed a new
approach based on nodes with one degree of freedom, a {\em memory}, indicating
the ability of the node to attract further links. We have found that with the
simple setting of the model the degree distribution converges to a power law,
where the exponent can be obtained analytically. As emergent properties of the
model, (1) preferential attachment is obtained, a feature observed recently in
various real growing networks, and (2) the correlation between age and linking
probability is negative, in agreement also with the empirical results mentioned
above. Unlike previous models, degree and age of nodes are uncorrelated in the
model introduced here.  Therefore the networks retain the power-law
distribution of the degree even though only the most recent nodes are
considered. This agrees with the fact that also truncated real-world networks
are observed to be scale-free. Finally it is worth noting the resemblance of
the grown networks to regular lattices.  The highly clustered scale-free
networks make a connection between scale-free networks and regular lattices.
They define a new class of scale-free networks. Interesting extensions of the
model include the introduction of random links, similarly to models of
small-world networks. We expect to find a connection between
scale-free growing networks and the small-world transition from regular
lattices. Research along this line is in progress.

\section*{Acknowledgements}
We would like to thank Anthony F.J.\ van Raan for providing us with the
age-distribution data in Figure \ref{fcitation}. VME acknowledges financial
support from the Danish Natural Science Research Council. We are grateful
to Emilio Hern\'andez-Garc\'\i a and Kristian Schaadt for useful comments on
the manuscript.

\ignore{
\clearpage
\begin{figure}
\centerline{\epsfig{file=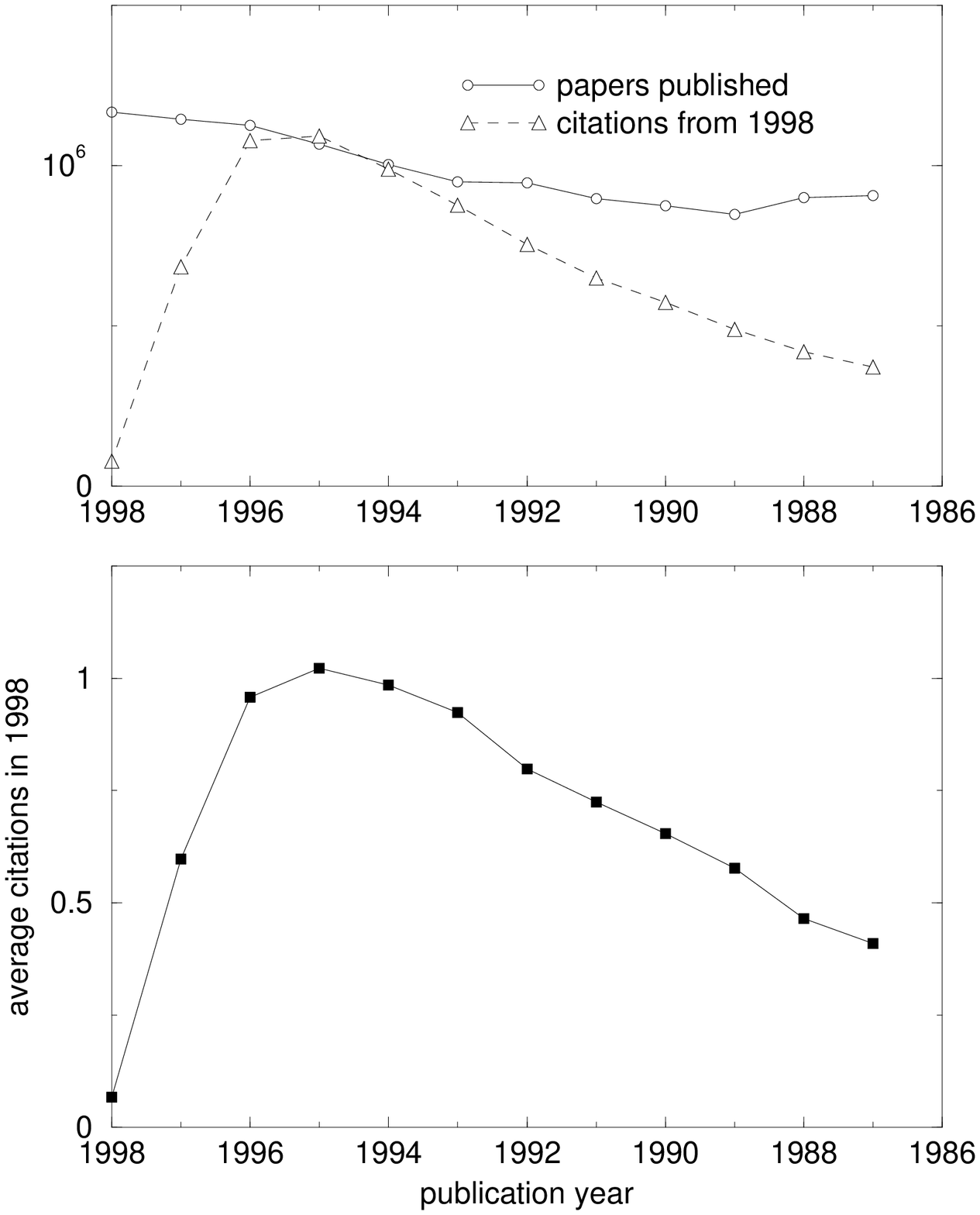,width=.6\textwidth}}
\end{figure}

\clearpage
\begin{figure}
\centerline{\epsfig{file=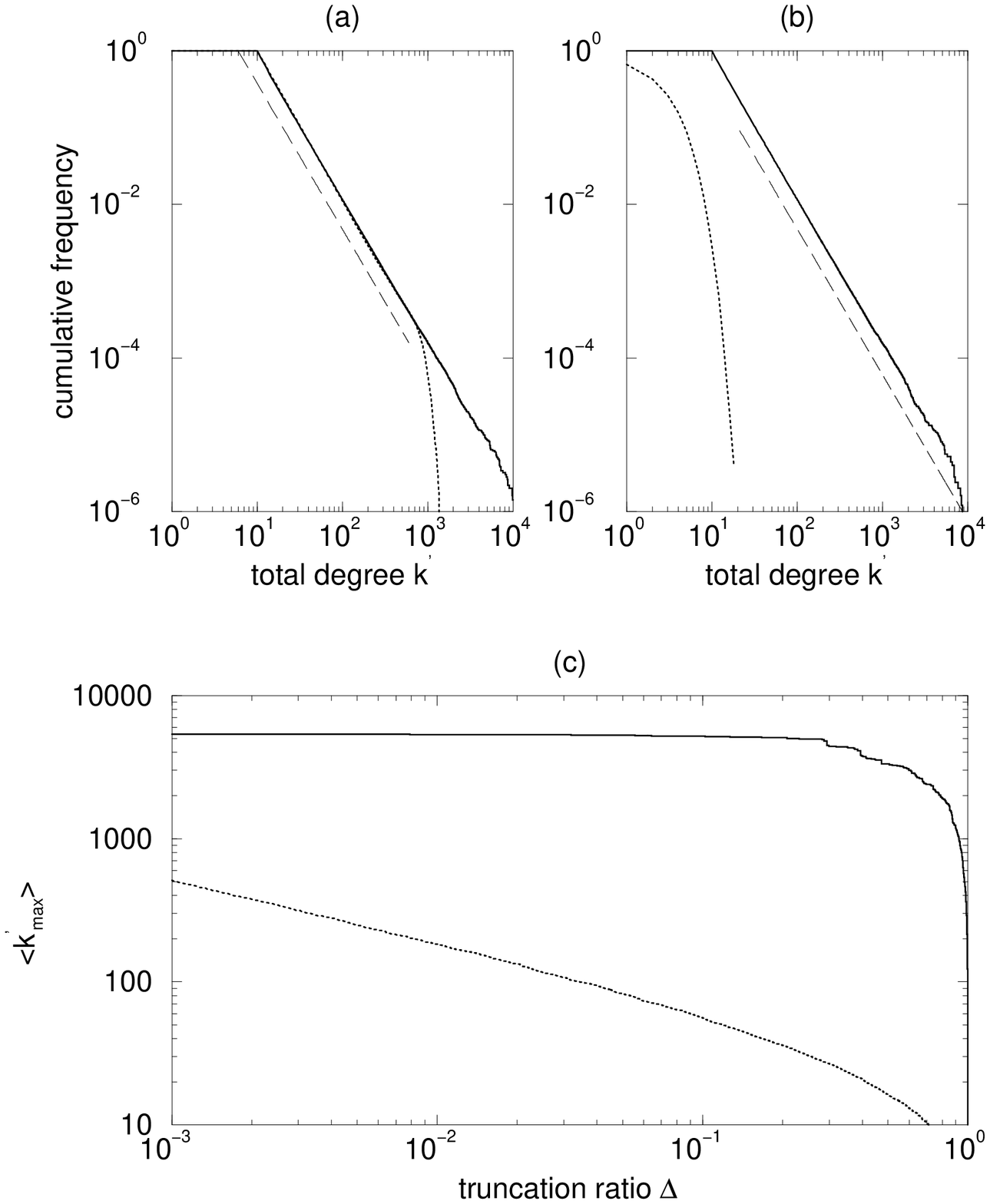,width=.6\textwidth}}
\end{figure}

\clearpage
\begin{figure}
\centerline{\epsfig{file=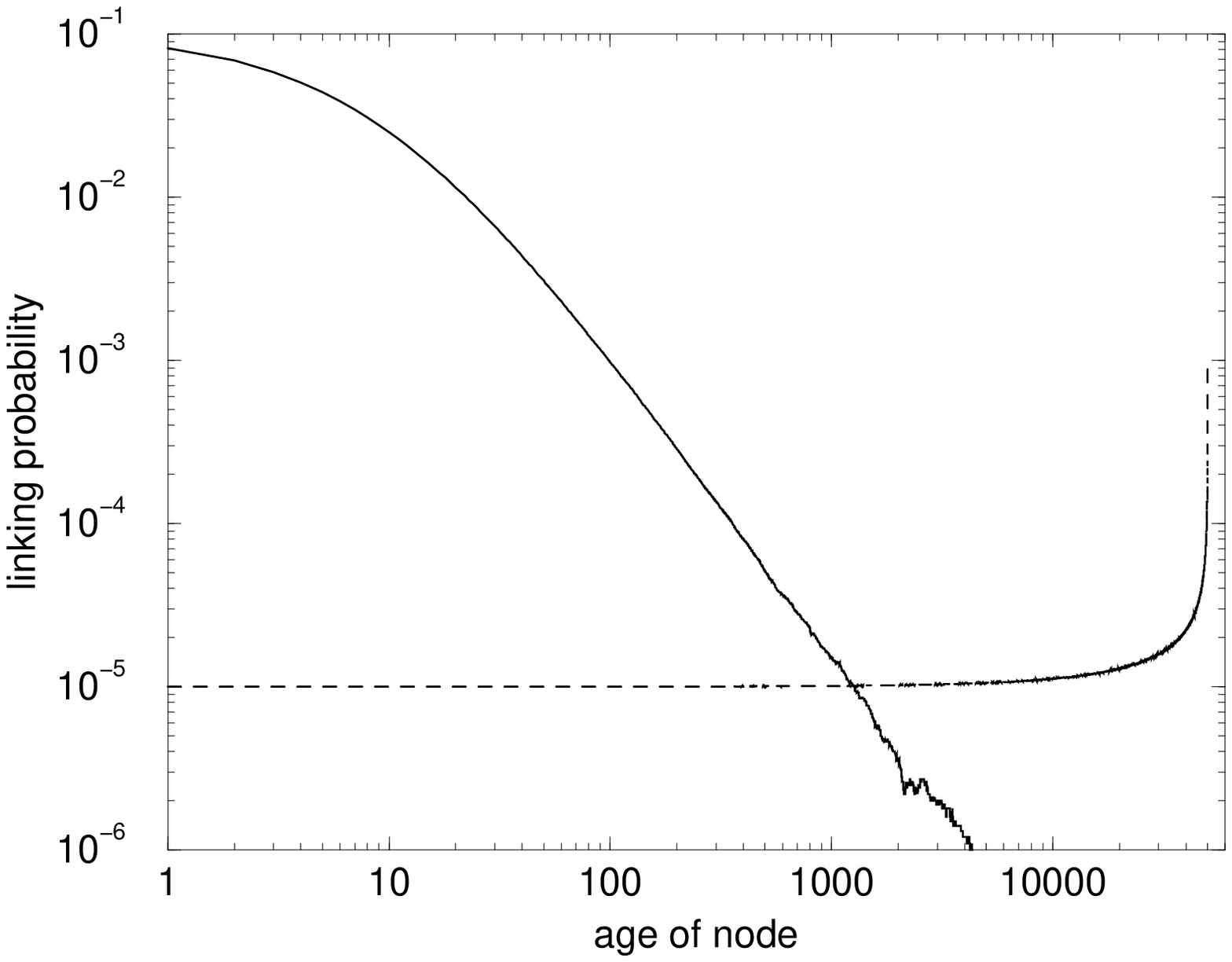,width=.6\textwidth}}
\end{figure}

\clearpage
\begin{figure}
\centerline{\epsfig{file=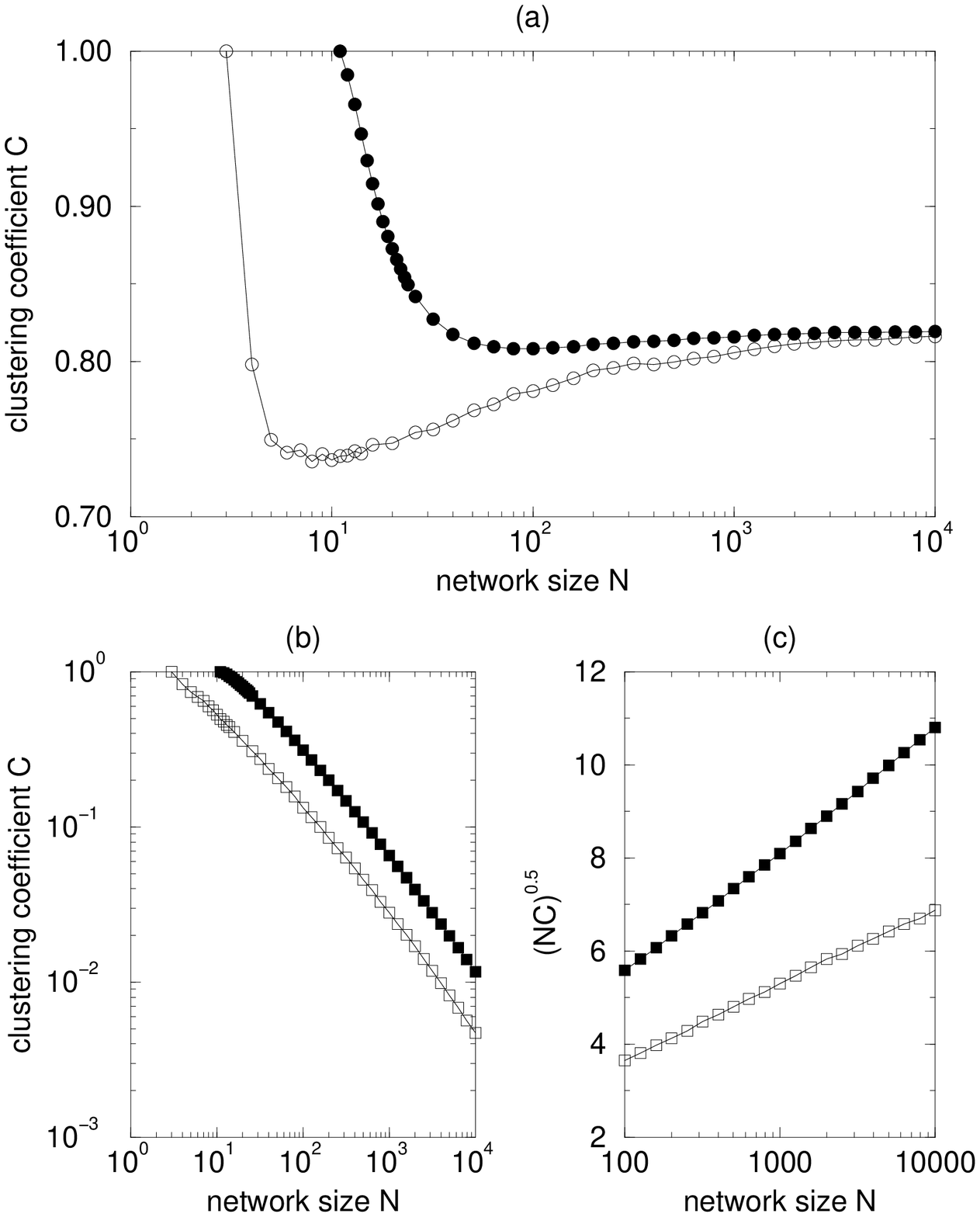,width=.6\textwidth}}
\end{figure}
}

\end{document}